\documentclass[letter,12pt]{article}
\usepackage{graphicx,amssymb,amsmath,epstopdf,color,hyperref}

\def\beq{\begin{equation}}
\def\eeq{\end{equation}}
\def\bea{\begin{eqnarray}}
\def\eea{\end{eqnarray}}

\def\gev{\, {\rm GeV}}

\newcommand{\gsim}{\lower.7ex\hbox{$\;\stackrel{\textstyle>}{\sim}\;$}}
\newcommand{\lsim}{\lower.7ex\hbox{$\;\stackrel{\textstyle<}{\sim}\;$}}

\def\DeltaFT{\Delta_{\rm FT}}

\def\xsection#1{\noindent{\bf #1}}

\begin{document}

\begin{flushright}
September 10, 2018
\end{flushright}

\vspace{0.07in}

\noindent
\begin{center}
{\bf\large Finetuned Cancellations and Improbable Theories}



\vspace{0.5cm}
{ James D. Wells}

{\it Leinweber Center for Theoretical Physics \\
Physics Department, University of Michigan \\
Ann Arbor, MI 48109-1040 USA}\\


\end{center}

\noindent
{\it Abstract:} 
It is argued that the $x-y$ cancellation model (XYCM) is a good proxy for discussions of finetuned cancellations in physical theories. XYCM is then analyzed from a  statistical perspective, where it is argued that  a finetuned point in the parameter space is not abnormal, with any such point being just as probable as any other point. 
However, landing inside a standardly defined finetuned region (i.e., the full parameter space of finetuned points) has  a much lower probability than landing outside the region, and that probability is invariant under assumed ranges of parameters. 
This proposition requires asserting also that the finetuned target region is  {\it a priori} established. Therefore, it is surmised that highly finetuned theories are generally expected to be highly improbable.   An actionable implication of this moderate naturalness position is that the search for a non-finetuned explanation to supplant an apparently finetuned theory is likely to be a valid pursuit,  but not guaranteed to be. A statistical characterization of this moderate position  is presented, as well as those of the extreme pro-naturalness and anti-naturalness positions.


\vfill\eject

\xsection{Introduction}

Assessing the theory quality~\cite{Wells:Lexicon} of naturalness of a theory has been an important and influential activity in particle physics for several decades~\cite{Giudice:2008bi,Fichet:2012sn,Farina:2013mla,Tavares:2013dga,Kawamura:2013kua,deGouvea:2014xba,Williams:2015gxa,Dine:2015xga,Wells:2016luz,Giudice:2017pzm,Hossenfelder:2018ikr,Wells:2018sus}. Work on this concept continues to the present day and is considered particularly critical since lack of new physics discoveries by the LHC to date puts the worth of naturalness and finetuning discussions under scrutiny. To remind the reader, when a theory point in the parameter space of a theory possesses Naturalness, or is ``natural", it is thought to be a point that is not unlikely. The algorithm to determine whether a theory is natural has often been deputized to a  finetuning functional~\cite{Ellis:1986yg,Barbieri:1987fn,Anderson:1994dz,Anderson:1994tr} the details of which will be discussed below. According to the standard interpretations a theory point that is highly finetuned is not natural and therefore considered improbable. It is this standard interpretation that will be addressed below. In particular, we introduce a simple algebraic model to make precise statements about the improbability of a theory landing within highly finetuned regions. 



Most discussions of finetuning, whether it be related to the Higgs boson mass, the $Z$ mass in supersymmetric theories, the dark matter relic abundance in supersymmetric theories, the cosmological constant, can be represented by  the cancellation of two parameters that results in a parameter with value much less than its antecedents. For example, finetuning in the Higgs boson discussion is often phrased as the Higgs bare mass-squared $m_0^2$ cancelling  a quantum correction quadratically sensitive to the cutoff of the effective theory $c\Lambda^2$, where $c$ is a coefficient not much smaller than $1$ and $\Lambda$ is the effective theory cutoff scale. The Standard Model Higgs mass is then $m^2_H=m_0^2+c\Lambda^2$. If the cutoff scale $\Lambda^2\gg m^2$  then a finetuned cancellation between $c\Lambda^2$ and $m_0^2$ must take place to yield $m_H^2\ll \Lambda^2$. It is thought that $\Lambda$ could take on any value up to the Planck scale $M_{Pl}\sim 10^{18}\gev$, which would require a finetuning to many decimal places to obtain $m_H\sim 10^2\gev$. 

In the above example, if we identify $c\Lambda^2/M_{Pl}^2$ with a variable $x$, $m_0^2/M_{Pl}^2$ with $-y$, and $m_H^2/M_{Pl}^2$ with $z$ we have the new equation $z=x-y$ where $x,y$ can take on values between 0 and 1, with $z$ required to be extremely small ($z\simeq 10^{-32}$ or so). This is an expression of the $x-y$ cancellation model (XYCM) for finetuning. A similar mapping to the XYCM is also a good representation\footnote{It is straightforward to generalize the probability discussions to come to models with more random variables cancelling. Nevertheless, the two-variable XYCM captures much of the essence of any multi-variable model.}  of the finetuning discussions of the cosmological constant problem, $Z$ mass in supersymmetry, etc.~\cite{Wells:2018sus}.

Among the several goals of this letter is to demonstrate a key tenet of the moderate naturalness position~\cite{Wells:2018sus}, which is the proposition that there is nothing special nor atypical with finetuned points in parameter space, and therefore they should not be viewed as  impossible or a monstrosity, yet landing in the finetuned region (space of all points exhibiting a finetuned cancellation) should {\it generally} be viewed as an improbable occurrence, and can rightly call into the question the completeness of a theory that requires a high finetuning to be empirically adequate. These propositions will be argued purely from statistical reasoning in the XYCM, without vague references to received wisdom in quantum field theory. Several additional implications of this moderate position will also be presented in the course of the discussion and in the conclusion.

\xsection{Probabilities in XYCM parameter space}

Let us begin by defining the XYCM in more detail, including providing the probability distributions of its parameters. As mentioned above, XYCM is the proposition that $z=x-y$ where $x$ and $y$ are independently and flatly distributed from 0 to 1 and $|z|\ll |x|,|y|$. The joint probability distribution function by the XYCM definition is $f(x,y)=1$ which satisfies the required unitary probability condition
\beq
1=\int_0^1dy\int_0^1dx\, f(x,y).
\eeq

\begin{figure}[t] 
\begin{center}
\includegraphics[width=0.6\textwidth]{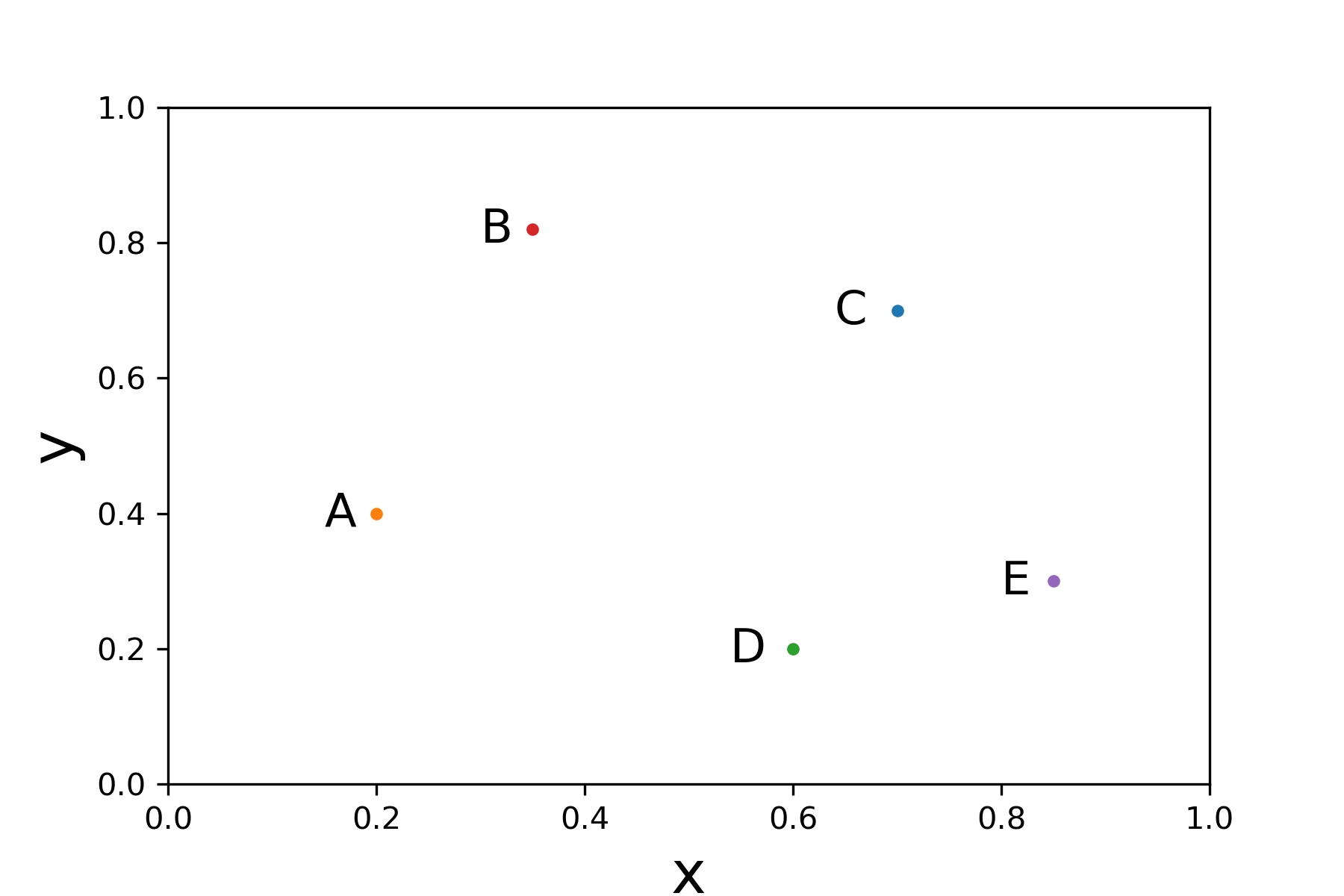} 
\caption{Five randomly sampled points on the flatly distributed $x$ and $y$ independent random variables. Each point is equally likely and there is nothing special or abnormal among them. However, point C is by definition a ``finetuned" point for $z$ when considering $z=x-y$ since it lies very near the $x=y$ diagonal.}
\label{fig:ax3}
\end{center}
\end{figure}

In Fig.~\ref{fig:ax3} the $(x,y)$ plane is shown with a few points (A,B,C,D,E) sampled according to the flat joint probability density function $f(x,y)=1$ over $0\leq x \leq 1$ and $0\leq y\leq 1$. One may ask if any of these points looks particularly unlikely or outrageously unexpected compared to any other? The answer is no, they are all equally likely. However, point C happens to lie very close to the diagonal characterized by $y\simeq x$.  As we shall see below, such a point is label ``finetuned" under a standard finetuning measure used in physics, and as such is sometimes considered difficult (``large finetuning") or impossible (``extreme finetuning") to contemplate by some. However, from a statistics point of view, such a point is by no means atypical, abnormal or monstrous. It is just as likely as any other point in the plot, and should have no disqualifying prejudices against it from a point-by-point perspective.

\xsection{Target regions and probabilities}

\begin{figure}[t] 
\begin{center}
\includegraphics[width=0.6\textwidth]{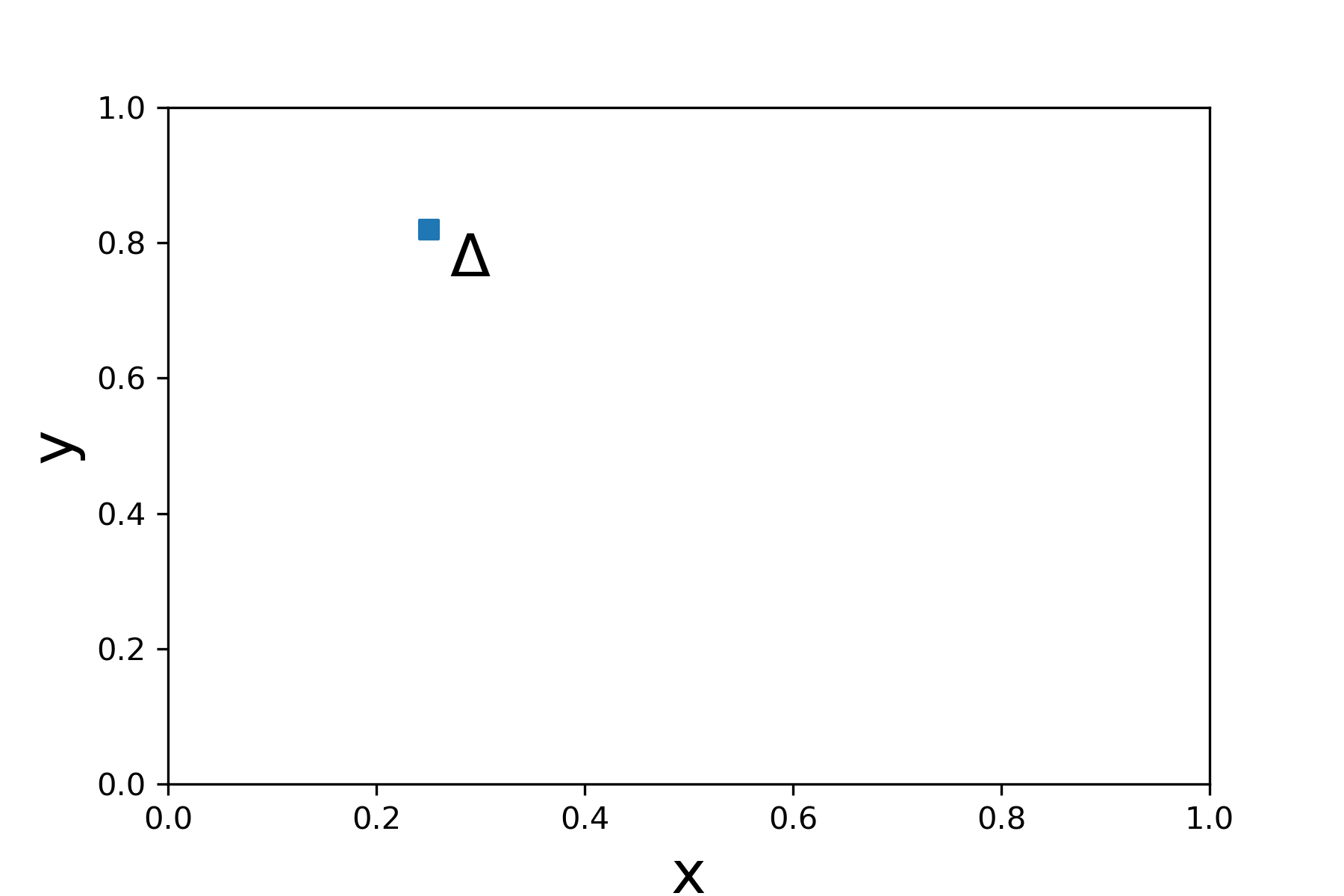} 
\caption{The small blue box $\Delta$ is arbitrarily placed in the the $xy$ plane. The probability of a subsequent single sampled point on the flatly distributed independent variables $x$ and $y$ to land within $\Delta$ is very small, and equal to $P_\Delta=V_\Delta$ where $V_\Delta$ is the area of the square.}
\label{fig:ax4}
\end{center}
\end{figure}

Now, suppose we identify a very small finite ``target region" in the $(x,y)$ plane. Let's call it $\Delta$, as depicted in Fig.~\ref{fig:ax4}. One can ask, what is the probability of sampling $f(x,y)$ and finding a point within $\Delta$? This is a very well defined question that can be answered. Bets can be cast intelligently on such questions. If the volume of $\Delta$ is $V_\Delta$ then, the probability $P_\Delta$ of a sampled point landing within $\Delta$ is $P_\Delta=V_\Delta$.

From a purely statistical point of view we can ask and answer an infinite number of similar questions based on different choices of the target region $\Delta$. The small regions do not have to be small boxes such as $\Delta$ in Fig.~\ref{fig:ax4}. One could also declare three other very small regions. Three choices out of an infinite number of choices are the region of very small values of $x<\epsilon$ but any value of $y$ ($\Delta_1$), or a very thin region where $|x+2y|<\epsilon$ ($\Delta_2$), or  a very thin region where $|x-y|\simeq \epsilon$ ($\Delta_3$). Each of these regions has volume $V_{\Delta_i}\simeq \epsilon\ll 1$. In each of these cases one can ask, what is the probability $P_{\Delta_i}$ of a sampled point landing within the region $\Delta_i$? The answer is $P_{\Delta_i}=V_{\Delta_i}\ll 1$, which is a very unlikely probability just as it is a very unlikely probability of sampling a single point within $\Delta$ in Fig.~\ref{fig:ax4}. 

\xsection{A priori vs.\ a posteriori statistical propositions}

An incorrect way of asserting probable versus improbable statistical propositions is to sample the distribution $f(x,y)$ for a point $(x_0,y_0)$, then draw an infinitesimally tiny box $\Delta_0$ around that point in the $(x,y)$ plane, and then declare that the sampled point that landed within tiny $\Delta_0$ was an incredibly improbable occurrence ($P_0=V_{\Delta_0}\ll 1$) that just should never have happened. No, such {\it a posteriori} reasoning is incorrect. In other words, you will never get somebody to take a bet that the sampled point will never fall into an arbitrarily tiny box you are free to draw {\it after} the point has been identified. 

On the other hand, going the opposite direction -- {\it a priori} reasoning -- does lead to valid statistical propositions of probable vs.\ improbable. This method entails identifying a tiny box $\Delta$ in the $(x,y)$ plane first and {\it then} sampling the $f(x,y)$ distribution for a point $(x_0,y_0)$. If $V_\Delta\ll 1$ then asking for the probability $P$ of $(x_0,y_0)$ landing within $\Delta$ has meaning and the answer is $P=V_\Delta$ within the XYCM, which is extraordinarily small.
 If such an occurrence  did happen it would  be cause for noting or commenting on being lucky or unlucky. However, it would not necessarily be sufficient evidence to disprove the $z=x-y$ model with its assumptions of independently distributed $x$ and $y$. The smaller the region $\Delta$ the more unlikely such a chance occurrence could have happened, and perhaps there is a sufficiently small target region $\Delta$ wherein one would not countenance a chance landing.  Of course, one must keep in mind that there is no violation of mathematics, physics, statistics or nature if $(x_0,y_0)$ were to fall within the  target region $\Delta$ no matter how small $V_\Delta$ is. However, admitting that nothing breaks if a highly improbable event happens should not give one license to ignore the fact that such occurrences are highly improbable and should generally not happen.
  
 
 \xsection{Analogy with gaussian distributions}
 
\begin{figure}[t] 
\begin{center}
\includegraphics[width=0.6\textwidth]{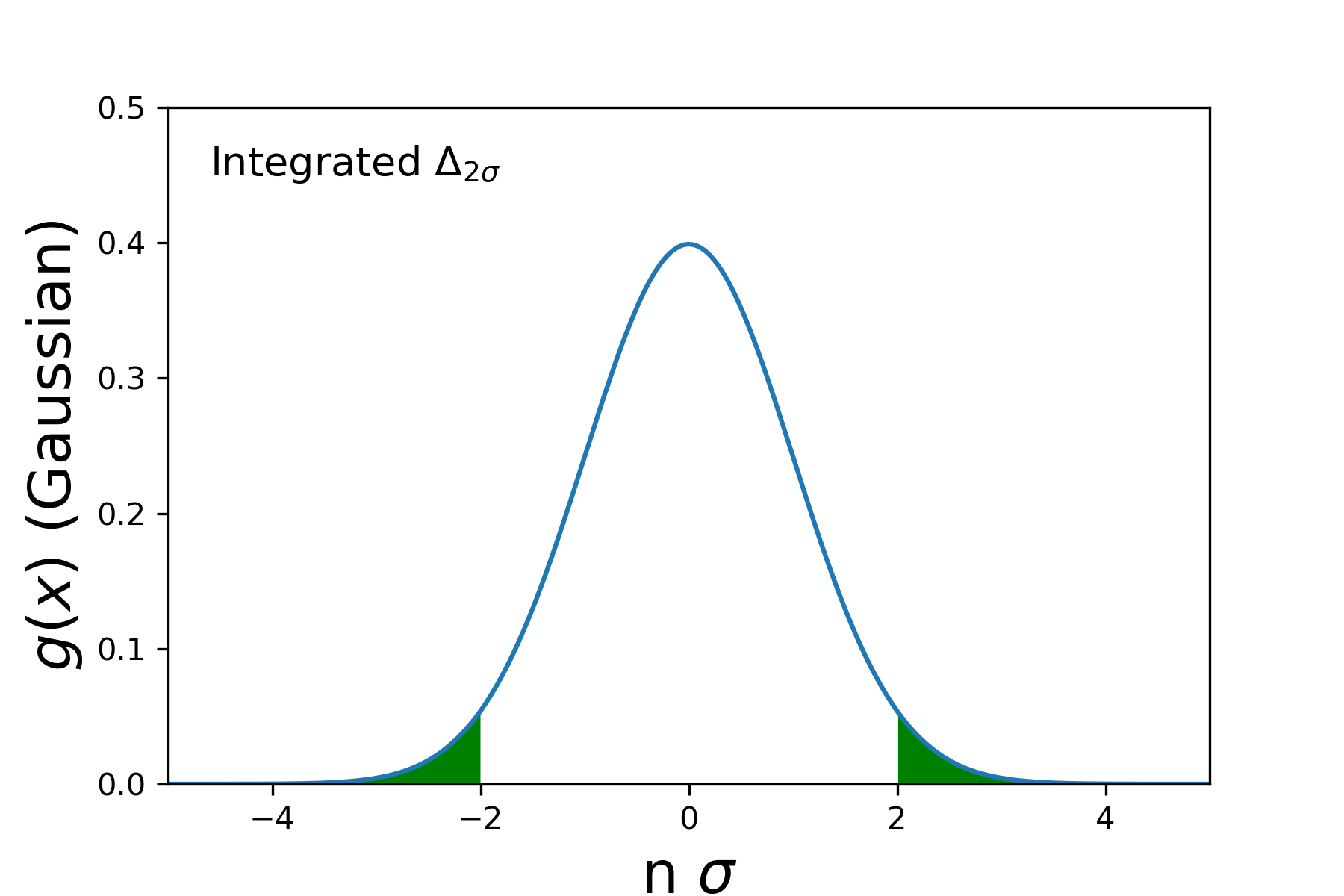} 
\caption{The green filled-in area is the target region $\Delta_{2\sigma}$ integrated over the gaussian distributed probability density function. It is rare that a randomly sampled single point should fall within $\Delta_{2\sigma}$ and rarer yet for it to fall within $\Delta_{3\sigma}$ (i.e., $|x|>3$), and so on.}
\label{fig:ax6}
\end{center}
\end{figure}

 The discussion above regarding probability assessments and {\it a priori} determined target regions is analogous to assessing probabilities of being more than a few standard deviations $\sigma$ away from the mean in a Gaussian distributed probability distribution centered at zero with unit standard deviation, 
 \beq
 g(x)=\frac{1}{\sqrt{2\pi}}\, e^{-x^2/2}.
 \eeq
 Declaring ``more than $2\sigma$ away" is equivalent to an {\it a priori} designation of a target region $\Delta_{2\sigma}$ (see Fig.~\ref{fig:ax6}) in parameter space of a random variable that is more than $2$ standard deviations away from the mean value. $\Delta_{2\sigma}$ is the region of $x$ where $|x|\geq 2$. Landing more than $2\sigma$ away is  a somewhat rare event, and it is equivalent to saying that the point landed in the $\Delta_{2\sigma}$ region which has small probability when integrated over $g(x)$. Landing within the $\Delta_{3\sigma}$ region is significantly more rare, and $\Delta_{5\sigma}$ rarer yet, etc.
 
Continuing with this analogy,  it is hard to say what measure-zero point is likely and what is not (landing on any true point is infinitesimally unlikely, yet it must land somewhere when sampled), but one can assess probabilities of being in a finite region, such as several $\sigma$ away from the mean. If just one sampling out of a handful is more than a few $\sigma$ away from the mean that is not cause for concern, but if there are several it means the theory has broken down. Likewise, if there is even one point many $\sigma$ away from the mean it is enough to declare the underlying theory broken. This is the standard way of expressing limits and discoveries in particle physics, such as ``$3\sigma$ discovery thresholds"~\cite{Conway:1999}.  The analogy here is to translate the above discussions into pure probabilities and assess how likely an occurrence is to happen. One sampled point found in an extraordinarily tiny target region $\Delta$ identified beforehand, or several points in a larger target region, both indicate that the underlying theory assumptions are likely to be wrong.

\xsection{Finetuning is just another {\it a priori} small target region}

Now we come to how this is related to finetuning in physics theories. The connection is that computing the highly finetuned region of parameter space for a theory is equivalent to identifying a small region $\Delta$, which we can call $\DeltaFT$. From a statistical point of view there is nothing special about a small region $\DeltaFT$ compared to any of the other small regions that one could identify. Thus, no points within the finetuned region are abnormal or special or strange -- they just happen to lie in a small region in parameter space. A key question is whether in the practical application of finetuning  to determine if a theory is natural or unnatural (i.e., likely or unlikely) one has identified $\DeltaFT$  as  an {\it a posteriori} selection or as an {\it a priori} selection. If {\it a posteriori} then declarations of probable versus improbable finetuned theory points are meaningless, but if {\it a priori} such declarations are meaningful. It is assumed in this letter that $\DeltaFT$ is an {\it a priori} determined region since it is algorithmically computed without reference to the details of a theory, and therefore probability reasoning is valid\footnote{An attack on this viewpoint (e.g., concern for what observables, or functions of observables, should be used to compute finetunings) and a defense against such attacks can be found in~\cite{Wells:2018sus}.}.

\begin{figure}[t] 
\begin{center}
\includegraphics[width=0.45\textwidth]{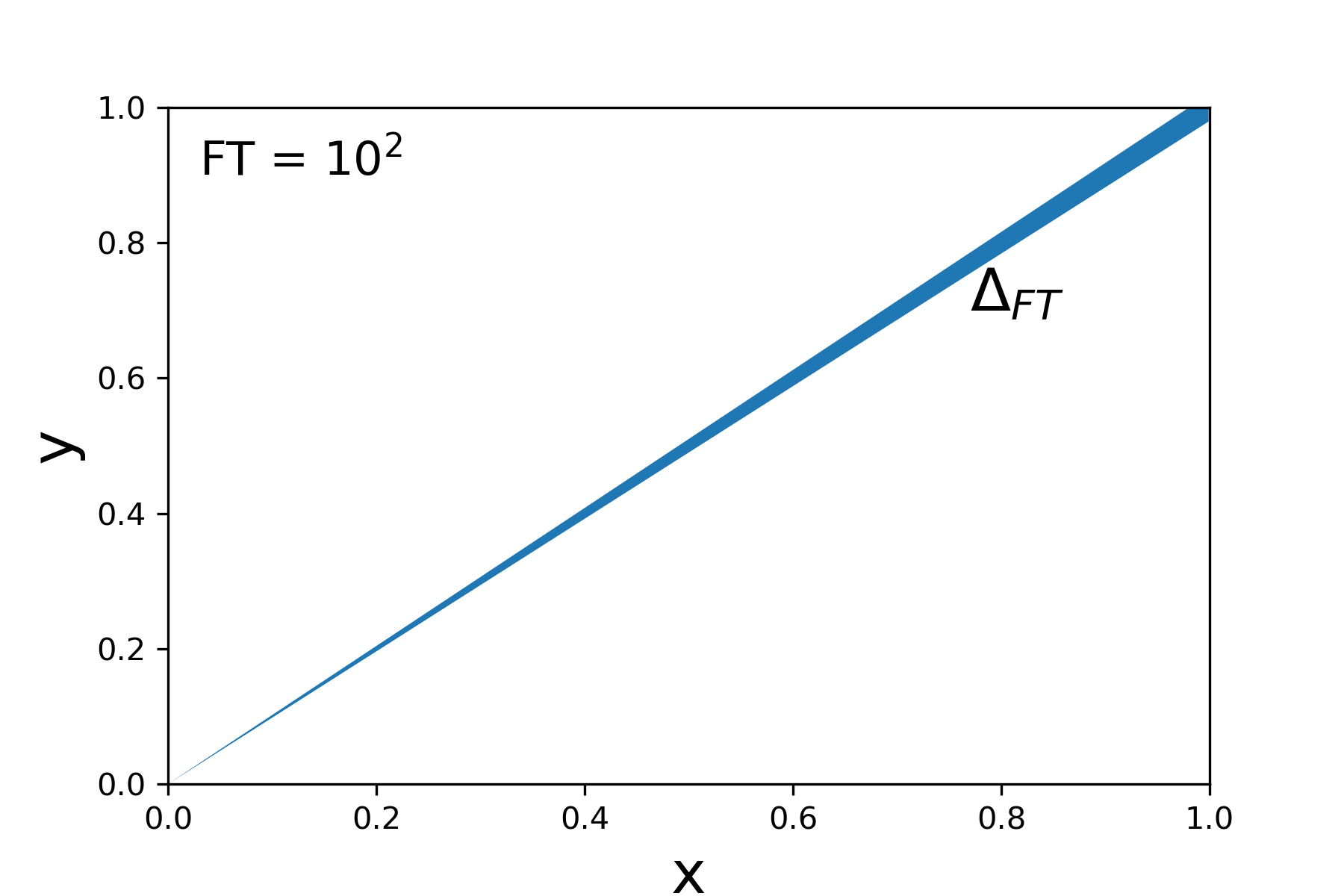} 
\includegraphics[width=0.45\textwidth]{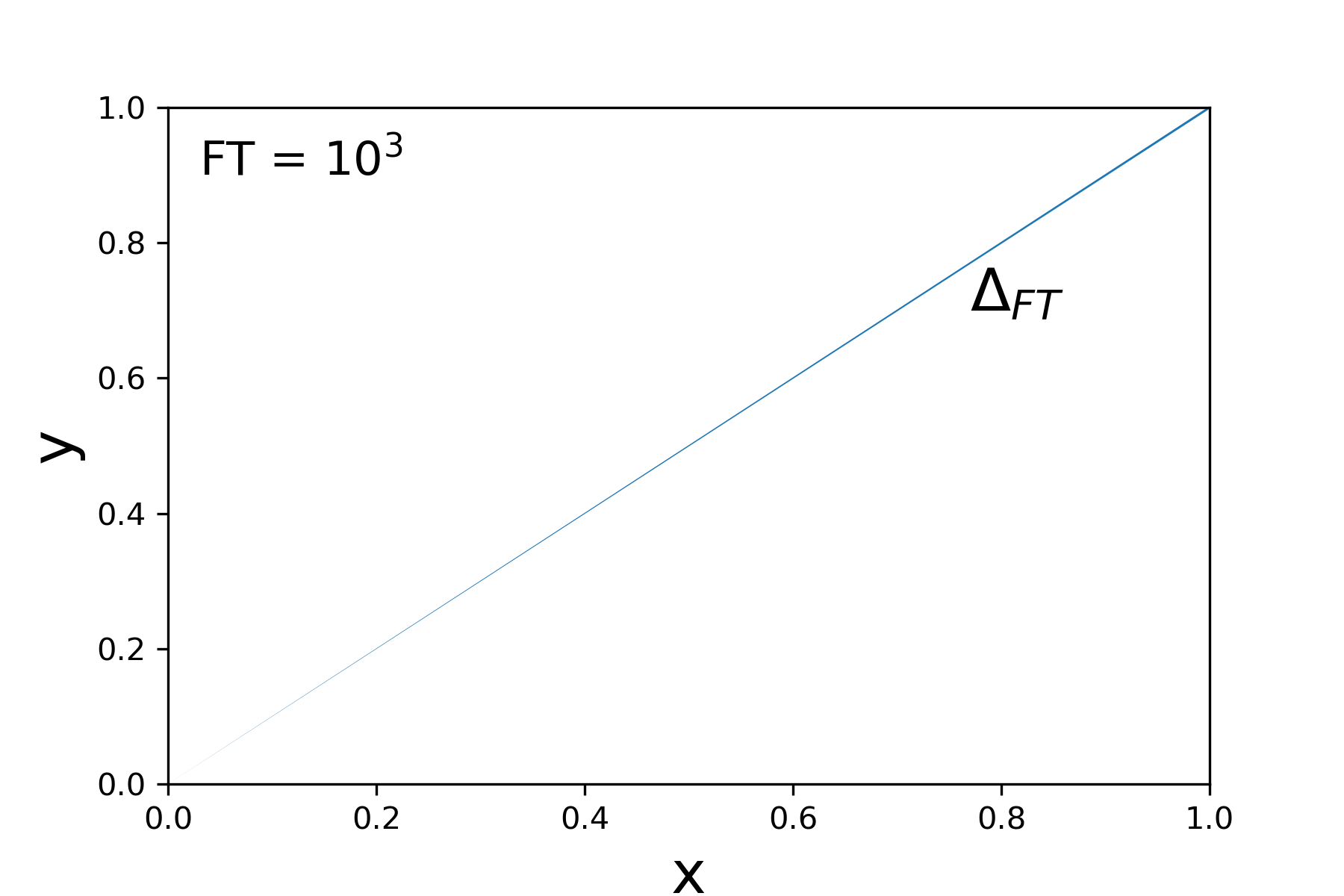} 
\caption{Plot of the region with $z$ finetuned according to eq.~\ref{eq:FT}, where $z=x-y$ and $x$ and $y$ are flatly distributed independent random variables from $0\leq x,y\leq 1$. {\it Left panel:} the blue filled-in region is where ${\rm FT}>10^2$. {\it Right panel:} the blue filled-in region is where ${\rm FT}>10^3$.}
\label{fig:ax2}
\end{center}
\end{figure}

Let us now show how finetuning selects a small region of parameter space. The common definition of finetuning of the parameters $x_i$ on some observable or outcome $\xi$ through the relation $\xi=\xi(x_1,x_2,\ldots,x_n)$ is
\beq
FT=\sum_{i=1}^n \left| \frac{x_i}{\xi}\frac{\partial \xi}{\partial x_i}\right|.
\eeq
In our XYCM this translates to
\beq
FT=\left|\frac{x}{z}\frac{\partial z}{\partial x}\right|+\left|\frac{y}{z}\frac{\partial z}{\partial y}\right|
=\frac{x+y}{|x-y|}
\label{eq:FT}
\eeq
A large finetuning is one where ${\rm FT}>10^2$ or $10^3$, for example. An extreme finetuning might be defined to be ${\rm FT}>10^6$. We plot the finetuned regions corresponding to ${\rm FT}>10^2$ and ${\rm FT}>10^3$ in Fig.~\ref{fig:ax2} (the region ${\rm FT}>10^6$ is too small to be visible). The target region volume corresponding to finetuning FT is
\beq
V_{\Delta_{\rm FT}}=\frac{2}{{\rm FT}+1}.
\eeq
For large ${\rm FT}$ the volume is very small. Therefore, it is improbable\footnote{If we had instead chosen $x,y$ to be flat over the interval $\xi\leq x,y\leq1$ instead of $0\leq x,y\leq 1$ the probability of landing in the FT region would increase  modestly to $P_{\DeltaFT}(\xi)=\frac{2}{{\rm FT}+1}\left(\frac{1+\xi}{1-\xi}\right)$ when ${\rm FT}\cdot (1-\xi)\gg 1$.} ($P_{\DeltaFT}=V_{\rm FT}$) that a sampled point on a flat distribution, or a nearly flat distribution spread out in $(x,y)$ on a scale much larger than $\DeltaFT$, should land in $\DeltaFT$. It is nigh impossible for it to land in the extreme finetuned region of ${\rm FT}>10^6$.  

\xsection{Finetunings and probable natural theories}

It is important to keep in mind that the finetuning computation, as normally defined (see eq.~\ref{eq:FT}), does not care about what the probability distribution is. It is an unambiguous definition. On the other hand, connecting finetuning to a probability requires integrating a probability density function $f(x,y)$ over a selected target region $\Delta$. Likewise, determining whether a sample point is improbable to land within the finetuned region requires integrating a probability density function over $\DeltaFT$, although making a qualitative binary declaration of  ``improbable" vs.\ ``not necessarily improbable" does not  require knowing the probability density function {\it precisely}. If it is flat or not too radically different from flat, then it is rare to fall into $\DeltaFT$. If it is peaked heavily along the diagonal, or in the upper right corner of the $xy$ plane, then landing in $\DeltaFT$ would be less improbable.

\begin{figure}[t] 
\begin{center}
\includegraphics[width=0.6\textwidth]{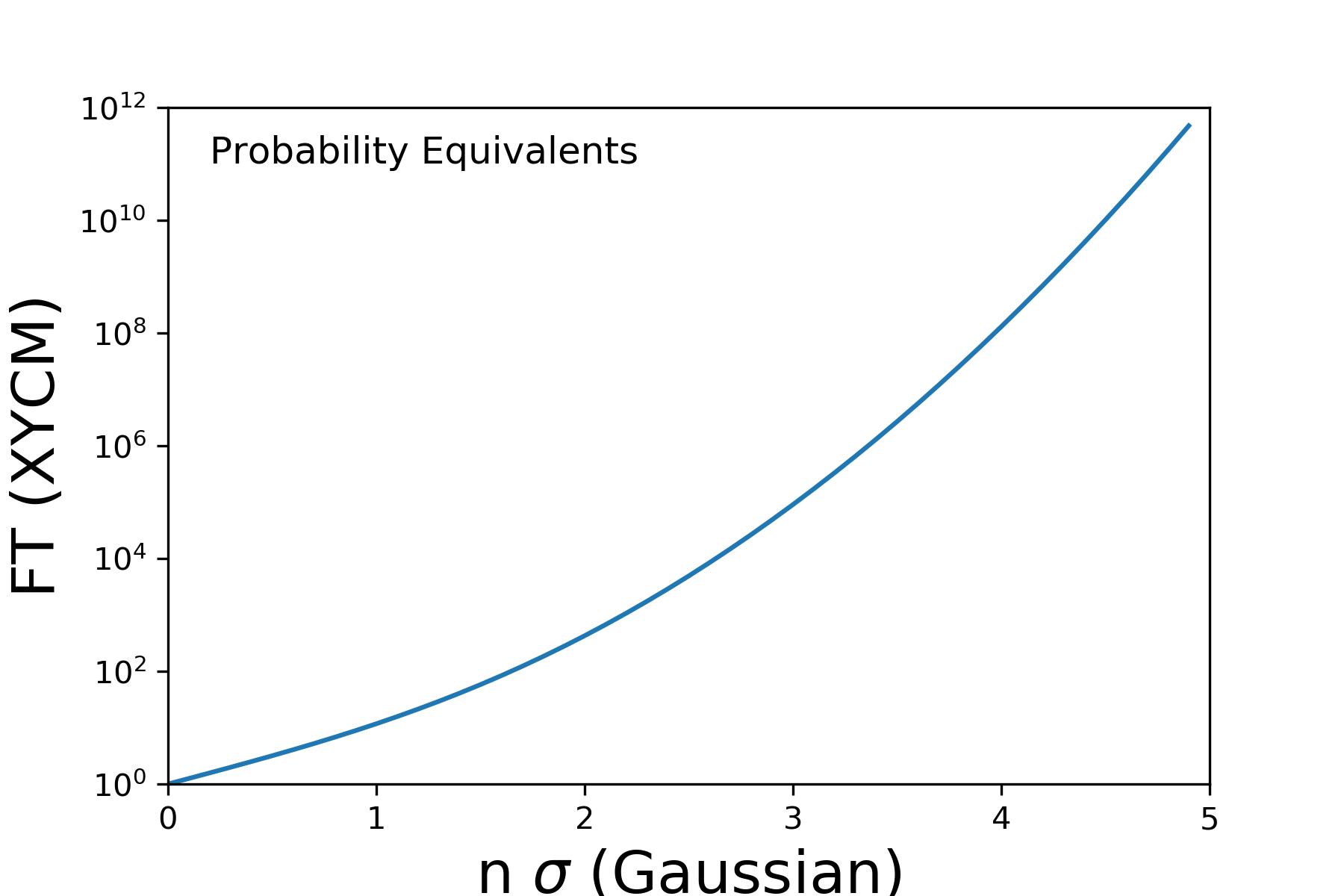} 
\caption{The line indicates where the probability of a single random variable sampled on a gaussian distribution landing more than $n\,\sigma$ away from its mean value is equal to the probability of a randomly selected single point in the flatly distributed $(x,y)$ plane in the XYCM landing within the region where finetuning is greater than FT.}
\label{fig:ax5}
\end{center}
\end{figure}

Back to the main discussion: The analogy with the statistical analysis of a single random variable gaussian distributed can be revisited. We can compute the FT value that would give the same probability of $(x,y)$ landing within $\DeltaFT$ as the probability of a single random variable $x$ landing within $\Delta_{n\sigma}$, which is the region of $x$ more than $n\sigma$ away from its mean value. One sees in Fig.~\ref{fig:ax5} that with increasing $n\sigma$ the FT increases rapidly. A ``$2\sigma$ event" has the same probability as a finetuning of more than $430$.  A ``$3\sigma$ event", which is certainly quite rare (probability of 0.27\%), is equivalent to finetuning of  more than $9\cdot 10^4$.  Given that $3\sigma$ signifies an important threshold of rarity and signs of new physics in physics\footnote{The importance of $3\sigma$ (99.73\% probability of falling within it) is highlighted in the Intergovernmental Panel on Climate Change (IPCC) Reports which give $>99\%$ likelihood the English phrase ``virtually certain"~\cite{IPCC}. Thus, it is ``virtually certain" that a sampled point should lie within $3\sigma$ of the mean.}, it would be appropriate to consider ${\rm FT}>10^6$ as an extreme finetuning that should not happen given standard assumptions.

One of the reasons that experiment requires very high ``$\sigma$" ($3\sigma$, or more\footnote{By ``$n\, \sigma$" one means that the measurement is more than $n$ standard deviations away from expected measurement within the standard theory under consideration. For very large $n$, the probability of the measurement fluctuation so far away from the true value is extremely rare, signifying a breakdown of the standard theory (i.e., breakdown of understanding of what physics is at play to give the results obtained).}) to declare that something new and interesting has taken place is because confidence in declaring that something new is happening beyond the standard theory is less sensitive to the assumption that the underlying measurement or theory uncertainty should be exactly gaussian~\cite{Cranmer:2015nia}, that it is plagued by small systematic effects, or that it suffers in significance due to the ``look elsewhere effect"~\cite{Gross:2010qma}. Likewise, it is expected that the higher the FT value that defines the boundary of the improbable region the less important it is to know very precisely the probability distribution to assume confidently that it is unlikely a sampled point will fall into it. This is a qualitative statement, since one can never rule out the possibility that the distribution is so highly peaked in the extreme finetuned region that a sampled point would indeed land in it. If that were the case, though, there would be a deeper principle yet to be articulated that explains that peaked distribution.

Here we bring in the concept of naturalness. A theory is considered natural if there is no reason to judge it to be improbable. If a theory, or rather theory point, is highly finetuned then it is considered improbable, and therefore is not natural\footnote{All the discussion in this article points more to the utility of declaring an unambiguous  threshold to decide if a theory is ``unnatural" (or improbable), since a theory on the other side of that threshold may also not be probable.}. For this reason declarations of natural or unnatural are often used interchangeable with declarations of finetuned or not finetuned, however we will continue to be careful here to declare a theory point finetuned or not based on the unambiguous definitions of the finetuning calculational procedure which may or may not be a precise guide to the probable, whereas natural and unnatural are reserved entirely based on probability assessments.

With this definition of naturalness in mind, 
the extreme anti-naturalness position~\cite{Hossenfelder:2018ikr} states that we know nothing about $f(x,y)$ and therefore should in no way ever attempt to even approximate what $P_{\DeltaFT}$ might be, and thus all statements that suggest that highly finetuned theories are improbable are nonsense and discussions of naturalness are useless. The extreme anti-naturalness position has many severe implications~\footnote{To name two specific examples, the extreme anti-naturalness position would imply that requiring an extreme splitting of the doublet and triplet in minimal grand unified theories~\cite{PDG:doublet-triplet} should not concern anyone, nor should we be concerned that the cosmological constant appears finetunely small compared to ordinary quantum field theory expectations~\cite{Weinberg:1988cp}.} that may be foreign to most practitioners in science as illustrated by the discussion of SEETA (skepticism against extra-empirical theory assessments) in~\cite{Wells:2018sus},  but the viewpoint cannot be fully dismissed since a meta theory of parameter distributions is not yet a developed concept for theories. Nevertheless, holding the extreme anti-naturalness position by reason of probability distributions not being exactly known looks  increasingly untenable as the threshold for acceptable FT increases (and $\DeltaFT$ decreases), just as the claims against paying attention to an $n\,\sigma$ signal in experiment by reason of measurements not being exactly gaussian distributed are increasingly untenable as $n$ increases.

The extreme pro-naturalness position, on the other hand, says that we can assume that $f(x,y)$ is close to flatly distributed (or some other precisely stated distribution) and so a large finetuning always means very low probability, and finetuning and naturalness are equivalent concepts for all practical purposes.  This perspective was argued to be dismissible in~\cite{Wells:2018sus} and there is no reason to revive it from the discussions here. 

The moderate naturalness position states that the range of parameters of $x$ and $y$ are generally on the order of their given values, that $f(x,y)$ is generally not radically different from flat over a large region that is much larger than and engulfs the highly finetuned region, and thus a large finetuning generally translates to a small probability. Naturalness and finetuning discussions are not interchangeable in the moderate position, but they are generally correlated well.

A mathematical characterization of the above descriptions is
\beq
P_{\DeltaFT}=\int_{\DeltaFT} dxdy\, f(x,y) = \left\{ \begin{array}{ll} {\rm impossible~to~estimate} & {\rm (extreme~anti\text{-}naturalness~position)} \\
V_{\DeltaFT} \ll 1&{\rm (extreme~pro\text{-}naturalness~position)} \\
\sim V_{\DeltaFT} \ll 1, ~{\rm generally} & {\rm (moderate~naturalness~position)}
\end{array}\right.\nonumber
\eeq
To emphasize, there is nothing controversial or unambiguous about the $\Delta_{\rm FT}$ part of the above equation, since it arises from a mere computational definition that returns the small $\Delta_{\rm FT}$ region. What is controversial, and is the origin of most differences of opinion, is what assumptions (explicitly stated or not) one makes for $f(x,y)$, as discussed above. In addition, the probability distributions can be significantly peaked by fixed-point renormalization group flows~\cite{Wells:2018sus} or by landscape statistics~\cite{Douglas:2003um}.

\xsection{Finetuning target region vs.\ small-$z$ target region}

Rather than defining the small region $\DeltaFT$ as a small target region as we did above we could have defined another very small target region in the $xy$ plane called $\Delta_z$ which is where $0<z<\epsilon$ (i.e., $0<x-y<\epsilon$) with $\epsilon \ll 1$, and then stated how unlikely it is that we should land in that region. There is no problem in doing that in the pure XYCM model, and the region can be {\it a priori} established and probability assessments made. This type of region is implicit among those who ask how probable it is that the Higgs boson should have been at its measured value or below, $P(0\leq m_H\leq m_H^{\rm expt})$. 

\begin{figure}[t] 
\begin{center}
\includegraphics[width=0.6\textwidth]{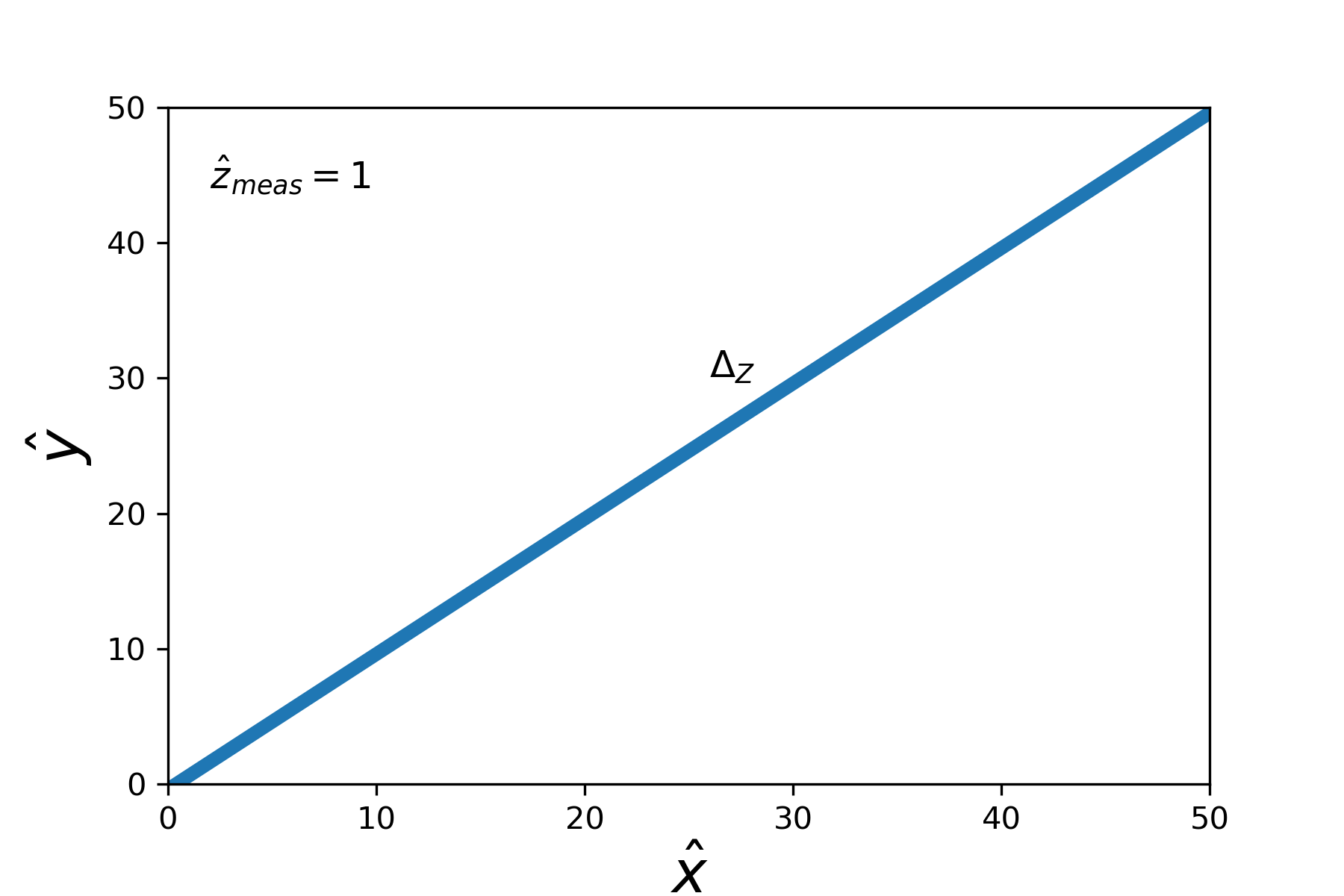} 
\caption{The filled-in blue region is the $\Delta_z$ target region in the rXYCM defined by $0<\hat z<\hat z_{\rm meas}$. Here we have normalized the measured value of $\hat z$ to be unity, $\hat z_{\rm meas}=1$.  We have also assumed the range $0\leq \hat x,\hat y\leq a$ with $a=50$ for this plot.}
\label{fig:ax7}
\end{center}
\end{figure}

However, there is a distinct advantage to preferring  $\DeltaFT$ over $\Delta_z$ when considering which {\it a priori} target region to connect to probability. For example, we often know $z$ --- let's call it $z_{\rm meas}$ --- but do not know $x$ and $y$, and we wish to know what values of $x$ and $y$ might be improbable. When knowing $z$ it is more convenient to  rescale our XYCM model such that $\hat z=\hat x-\hat y$, where $\hat z_{\rm meas}=1$.  We can call this the rXYCM\footnote{The rescaled variables of the rXYCM are $\hat x,\hat y,\hat z$ to not be mistaken with XYCM variables $x,y,z$.}  (rescaled XYCM) with $a$ the scale factor associated with the range on the rescaled variables $\hat x$ and $\hat y$ ($0\leq \hat x,\hat y\leq a$). 
The $\Delta_z$ region in the rXYCM is shown in Fig.~\ref{fig:ax7}.

\begin{figure}[t] 
\begin{center}
\includegraphics[width=0.6\textwidth]{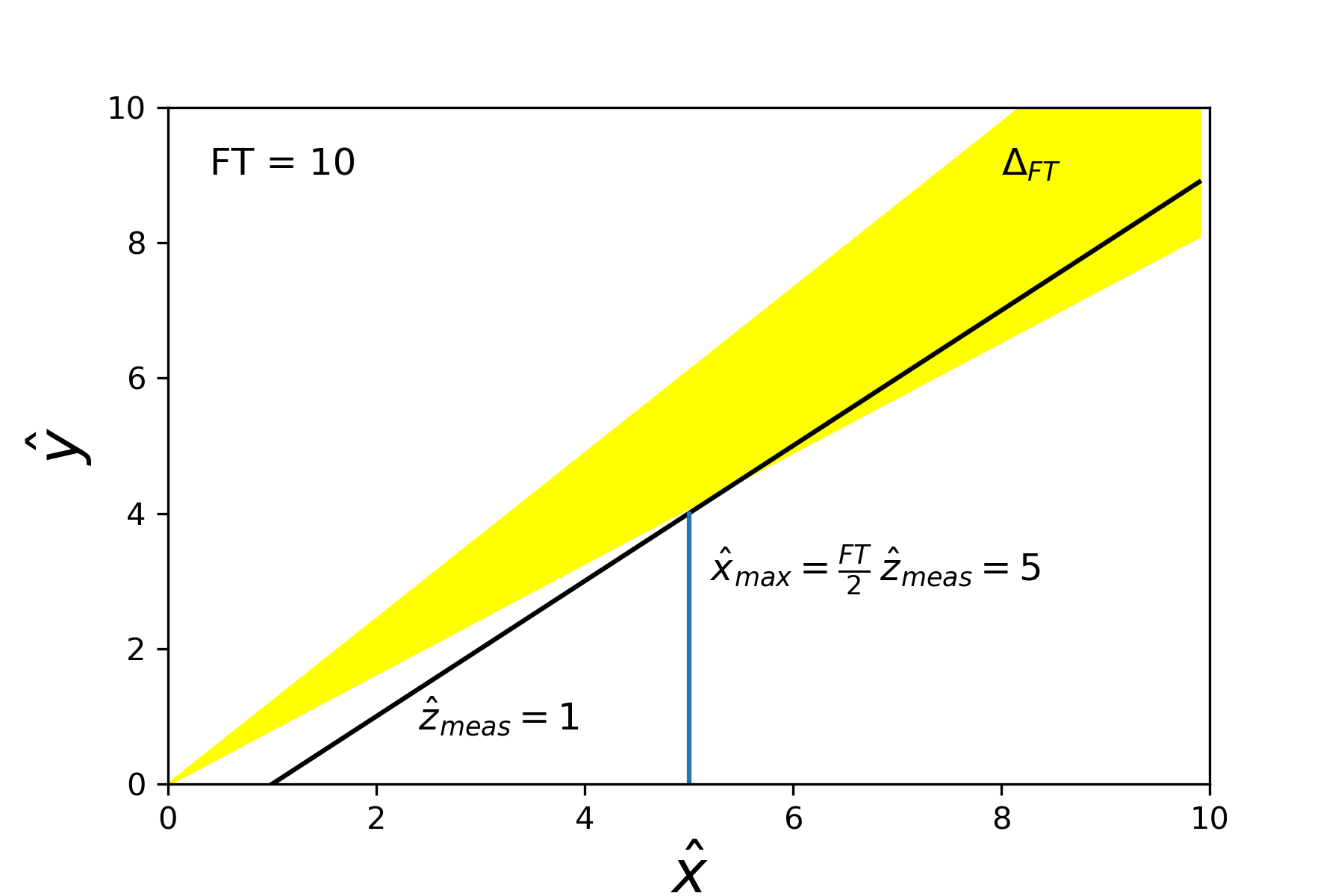} 
\caption{The filled-in yellow region is the $\DeltaFT$ target region in the rXYCM for ${\rm FT}=10$. This low value of ${\rm FT}=10$ is not advocated to be a good boundary for FT-improbable theory but is rather chosen here for more clear visual representation. The solid black line is the measured value of $\hat z$, which is normalized to unity here, $\hat z_{\rm meas}=1$. The maximum value that $\hat x$ can have along the $\hat z_{\rm meas}$ line without entering into the improbable $\DeltaFT$ region is $\hat x_{\rm max}=\frac{1}{2}{\rm FT}\cdot \hat z_{\rm meas}=5$. $\hat x_{\rm max}$ represents the upper bound on $\hat x,\hat y$ for the theory to not be FT-improbable. We have assumed the range $0\leq \hat x,\hat y\leq a$ with $a=10$ for this plot, but the value of $\hat x_{\rm max}$ is insensitive to $a$, which is partly what makes  $\DeltaFT$ a good choice for improbable target region analysis.}
\label{fig:ax8}
\end{center}
\end{figure}

In the $\Delta_z$  approach (note, $\Delta_z$ is where $0<\hat z<1$ in rXYCM) the probability of a randomly sampled point landing within $\Delta_z$  scales inversely with $a$ and is given by $P\sim \hat z_{\rm meas}/a$. Thus there is a  sensitivity to the assumed range of $\hat x$ and $\hat y$, and that is before any discussion about variations of probability density in the $\hat x\hat y$ plane.
On the other hand, in the $\DeltaFT$ analysis the probability of a randomly sampled point landing within $\DeltaFT$ remains $P=2/({\rm FT}+1)$ and is thus  independent of the scaling factor $a$. The values of $\hat x$ and $\hat y$ that are above $\frac{1}{2}{\rm FT}\cdot z_{\rm meas}$ are within $\DeltaFT$ and therefore improbable if FT is chosen large enough. See Fig.~\ref{fig:ax8} for a demonstration of this for ${\rm FT}=10$ and $\hat z_{\rm meas}=1$. The maximum values of $\hat x$ and $\hat y$ in this analysis are  insensitive to the scale factor $a$. This makes the $\DeltaFT$ analysis  more robust than a $\Delta_z$ analysis. In the next section we will use rXYCM and the preferred $\DeltaFT$ target region analysis to discuss how one might declare a speculative theory to be improbable.

\xsection{Declaring speculative theories improbable}

As described at the beginning many physicists consider the Standard Model to be a finetuned theory based on the implied cancellations needed for a small Higgs mass compared to the Planck scale. The moderate position on naturalness suggests that it is worthwhile to pursue a non-finetuned explanation of the Higgs sector, since it is likely to exist. It is not possible to declare a speculative new theory to be probable before confronting it with a suitable experimental effort, since there are presumably a large number of candidate new theories. However, it may be possible to declare it ``likely to be improbable" based on the finetuning assessments described above. Let's call a theory that suffers from finetuning to be ``FT-improbable."

Let's suppose that a speculative new theory has built into it a cancellation that can mapped to the rXYCM. For example, in a supersymmetric Standard Model (MSSM) the matching conditions between the full theory and the SM effective theory involves superpartner masses cancelling to yield the $Z$ mass~\cite{Martin:1997ns}:
\beq
\frac{m_Z^2}{2}+|\mu|^2=\frac{m_{H_d}^2-m_{H_u}^2\tan^2\beta}{\tan^2\beta-1}
\eeq
where $m^2_{H_u}$ and $m_{H_d}^2$ are the supersymmetric soft masses of the $H_u$ and $H_d$ Higgs bosons and $\tan\beta$ is the ratio of vacuum expectation values $\tan\beta=\langle H_u\rangle/\langle H_d\rangle$.
This maps to the rXYCM of the previous section by identifying
\beq
\hat x=\frac{m_{H_d^2}}{(m_Z^2/2)}\left(\frac{1}{\tan^2\beta-1}\right),~~\hat y=\frac{|\mu|^2}{(m_Z^2/2)}+\frac{m_{H_u}^2}{(m_Z^2/2)}\left(\frac{1}{\tan^2\beta-1}\right),~~\hat z_{\rm meas}=\frac{m_Z^2}{2}\frac{1}{(m_Z^2/2)}=1.\nonumber
\eeq
where $\hat z=z/(m_Z^2/2)$ and $\hat x$ and $\hat y$ are flatly distributed up to some unknown large value $a\sim \tilde m^2/m_Z^2$, with $\mu^2,m_{H_u}^2,m_{H_d}^2\sim \tilde m^2$ being typical supersymmetry mass scale. Again, it must be emphasized that it is unlikely that $\hat x$ and $\hat y$ are exactly flatly distributed in real physics theories for several reasons, including the fact that notions of distributions are scale dependent~\cite{Wells:2018sus}. However, we will assume that the distribution is generally not too radically different than flat and can be approximated well enough by flat.

If $\tilde m\gg m_Z$ then we are in a situation where  the speculative $\hat x$ and $\hat y$, which are distributed up to $a\sim \tilde m^2/m_Z^2 \gg 1$, must finetune themselves to cancel and obtain $\hat z_{\rm meas}=1$. Specifically, if $\hat x,\hat y$ are greater than ${\rm FT}/2$ a finetuning of FT between $\hat x$ and $\hat y$ is required to achieve $\hat z_{\rm meas}=1$, which is improbable because  landing within $\DeltaFT$ has a small probability when FT is large. By the arguments above it can be determined that large $m_{H_u},m_{H_d},|\mu|\gsim m_Z \sqrt{{\rm FT}}$ is improbable if the new speculative theory is correct. Thus,  very large superpartner masses (equivalently, $\hat x,\hat y\gg 1$ within rXYCM) imply an FT-improbable theory that is generally not expected to hold. Again, the converse is not valid -- the proposition that the theory is probable if not finetuned does not follow from the above discussion.

It could be the case that experiment has ruled out every speculative point in parameter space (i.e., the $\hat x$ and $\hat y$ superpartner mass values) that gives a non-finetuned result of a required observable (i.e., the measured $z$ observable).  In that case, the speculative theory is likely, but not guaranteed, to be wrong. Before making that assessment, however, one needs to determine that all non-finetuned points have indeed been ruled out by experiment, since premature judgments run the risk of  effectively expanding the target region larger than $\Delta_{\rm FT}$ to include the regions outside of $\DeltaFT$ that experiment cannot reach. Such an enhanced region perhaps could be {\it a priori} established as a tiny region $\DeltaFT'=\DeltaFT+\Delta_{X}$, where $\Delta_X$ is the region where experiment cannot reach but yet is not finetuned. It is safe to ignore $\Delta_X$ and declare a non-finetuned version of the speculative theory ruled out only if it can be established that $\Delta_X$ is a very tiny region\footnote{More technically, the probability density integral over $\Delta_X$ must yield a tiny probability for falling within $\Delta_X$.} of the full non-finetuned parameter space of the theory that is consistent with experiment. 

One may also invoke new theories that recast an XYCM apparent finetuning, such as the Standard Model, into a different structure that may, for example, involve exponential factors, such as theories of dimensional transmutation and theories of warped extra dimensions. In the warped extra-dimensions scenario the Higgs mass is obtained not entirely by an $x-y$ cancellation but also by a dimensional transmutation $z=(x-y)e^{-r/r_0}$.  If we assume flat distributions, or distributions not too dissimilar from flat, on $x$, $y$ and $r$ which are order unity, one finds that even very small values of $z\ll 1$ are not necessarily finetuned. In this case the theory is not FT-improbable. This is an example of a pursuit of a non-finetuned explanation (warping) for a theory (Higgs sector in Standard Model) that otherwise appeared to be finetuned. However, if experiment lowers the highest tolerated value of $r$ through searches of correlating phenomena (e.g., KK excitations of Standard Model states with ${\rm mass}\sim r^{-1}$), the cancellations between $x$ and $y$ needed to reach small $z$ may then be finetuned and the theory is FT-improbable. This may be the case currently in Randall-Sundrum theories of warped extra dimensions~\cite{Randall:1999ee}, although a careful and thorough technical analysis of its parameter space is required to make that claim.

\xsection{Summary}

It has been argued above that many finetuned cancellation discussions in physics can be mapped approximately to the XYCM. If distributions on parameters are reasonably independent and not too radically different from flat it is improbable that a theory point should land within a highly finetuned region. To justify this supposition we had to accept that finetuned regions are {\it a priori} defined regions, albeit not special or abnormal regions from a statistical distribution point of view, and thus it is statistically valid to make  probabilistic judgments on the likelihood of theory points falling within them. Furthermore, we have  carefully stated where some knowledge of the distribution of parameters $f(x,y)$ is required to connect a finetuned region, which is unambiguously determined, with (im)probability.

The moderate naturalness position, which the author is partial to, suggests that $f(x,y)$ is not too radically different than flat over a region larger than the finetuned region $\DeltaFT$, which allows one to declare that landing in the highly finetuned region is much less probable than landing outside of it. Nevertheless, the moderate naturalness position is tentative about the properties of the distribution $f(x,y)$ and is willing only to say that the previous sentence may be {\it generally true} but not guaranteed to be true. Implications of the moderate position are that searching for a non-finetuned explanation of what appears to be a finetuned empirically adequate theory is a valid enterprise, since there is a general expectation that a highly finetuned theory is highly improbable under a large range of conditions as argued above. Corollaries and re-phrasings of these conclusions are many, including the proposition that there is a high, but not guaranteed, probability that a non-finetuned new theory exists that can supplant an established theory that appears to be highly finetuned, and any speculative new theory that requires a high finetuning to conform with experiment is improbable.

\medskip\noindent
{\it Acknowledgments:}  This work was JDW supported in part by the DOE under grant DE-SC0007859. I wish to thank G. Giudice, S. Martin, A. Pierce, N. Steinberg and Y. Zhao for helpful conversations on these issues.


\end{document}